\documentclass[twocolumn,,showpacs,preprintnumbers,amsmath,amssymb,superscriptaddress]{revtex4}
\usepackage{graphicx} 
\usepackage{dcolumn} 
\usepackage{bm} 
\begin{document}

\preprint{APS/123-QED}

\title{Hole maximum density droplets of an antidot in strong magnetic fields }
\author{N. Y. Hwang}
\author{S.-R. Eric Yang\footnote{corresponding author,  eyang@venus.korea.ac.kr}}
\affiliation{Department of Physics, Korea University, Seoul 136-701, Korea} 
\author{H.-S. Sim}
\affiliation{School of Physics, Korea Institute for Advanced Study, Seoul 130-722, Korea}
\author{Hangmo Yi}
\affiliation{Department of Physics, Soongsil University, Seoul, 156-743, Korea}
\affiliation{School of Physics, Korea Institute for Advanced Study, Seoul 130-722, Korea}

 

\date{\today}

\begin{abstract}
We investigate a quantum antidot in the integer quantum Hall regime
(the filling factor is two) 
by using a Hartree-Fock approach and by transforming the electron antidot
into a system which confines holes via an electron-hole transformation.
We find that its ground state is the maximum density droplet of holes 
in certain parameter ranges. 
The competition between electron-electron interactions and the
confinement potential governs the properties of the hole droplet
such as its spin configuration.
The ground-state transitions between the droplets with different spin
configurations occur as magnetic field varies.
For a bell-shape antidot containing about 300 holes, 
the features of the transitions are in good agreement with
the predictions of a recently proposed capacitive interaction model
for antidots as well as recent experimental observations.
We show this agreement by obtaining
the parameters of the capacitive interaction model 
from the Hartree-Fock results.
An inverse parabolic antidot is also studied.
Its ground-state transitions, however, 
display different magnetic-field dependence from that of
a bell-shape antidot.
Our study demonstrates that the shape of antidot potential affects 
its physical properties significantly.

\end{abstract}


\maketitle

\section{Introduction}

\begin{figure}
\includegraphics[width=  \columnwidth]{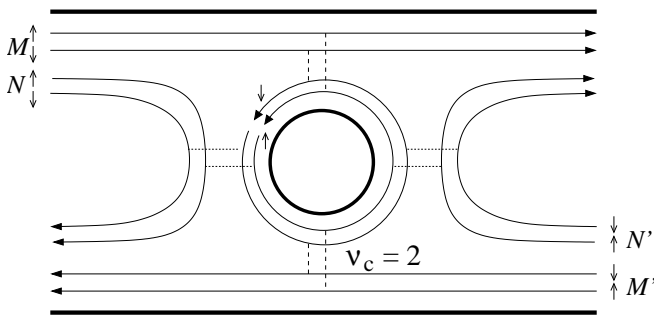}
\caption{\label{fig:experiment}
A quantum Hall antidot with localized antidot states and 
extended edge channels.
The localized states are weakly coupled to the extended channels,
as indicated by dashed and dotted line.
The small arrows show the spin direction of the states.
The local filling factor around the antidot is two.
}
\end{figure}

A quantum antidot has been extensively investigated experimentally
\cite{Hwang,Ford,Kirczenow2,Sachrajda,Goldman,Mace,Maasilta,
Masaya_charging,Masaya_double,Maasilta2,Karakurt,Masaya_Kondo}
and theoretically 
\cite{Kirczenow2,Mace,Kirczenow,Bogachek,Sim,antidot_PhysicaE} last decade.
It is a potential hill in two dimensional electron gas (2DEG) systems.
In zero magnetic field,
it is a simple repulsive potential and acts as a scattering center for
electrons.
In this sense, it is opposite to a quantum dot which confines electrons.
When a strong magnetic field is applied perpendicular to 2DEG,
the antidot has its own electronic ``edge'' structures, which  
correspond to classical skipping orbits around the antidot resulting
from the Lorentz force.
These localized antidot structures can be experimentally studied
by measuring conductance   
when they are weakly coupled to extended edge channels
propagating along the boundary of 2DEG.
In the integer quantum Hall regime,
the measured conductance exhibits
interesting Aharonov-Bohm oscillations
\cite{Ford,Sachrajda,Masaya_charging,Masaya_double,Masaya_Kondo},
which can not be understood within a single-particle picture.
For examples, the oscillations are accompanied by 
the charging effect \cite{Ford,Sachrajda,Masaya_charging},
nontrivial $h/(2e)$ Aharonov-Bohm 
oscillations \cite{Masaya_double,Masaya_Kondo},
and/or Kondo-like signatures \cite{Masaya_Kondo}.
These observations indicate that electron-electron interactions can
play an important role in the antidot system.

However,
there have been few theoretical works
on the interaction effects in the antidot system.
Very recently, a phenomenological capacitive interaction model has been proposed
\cite{Sim}
to explain the experimental results.
This model is based on 
the capacitive couplings between localized excess charges,
which are formed around the antidot due to magnetic flux quantization.
The capacitive interaction of the excess charges results in
Coulomb blockade and tunnelings are allowed only under certain 
conditions.  The main result of the above mentioned work was that 
the usual resonant tunnelings 
are accompanied by Kondo resonances, hence leading to nontrivial 
Aharonov-Bohm oscillations.
These predictions are in qualitatively good agreement with the
experimental observations \cite{Masaya_double,Masaya_Kondo}.
Based on a Hartree-Fock approach and a particle-hole transformation,
it was also suggested \cite{Sim,antidot_PhysicaE}
that holes inside an antidot can form
a maximum density droplet (MDD) \cite{Yang1,Yang2,Yang3,Oosterkamp}
in the ground state within some parameter ranges
and that the transitions between MDDs may lead to
Kondo effects,
supporting the capacitive interaction model.
However, the tested antidot was so small (it has about 50 holes) that
the transitions did not  occur  periodically with varying magnetic field,
in contrast to the experimental data.

In this paper, we develop a microscopic Hartree-Fock approach 
and apply it to a large-size antidot containing about 300 holes.  
Our approach is based on an electron-hole transformation,
where an antidot potential of electrons is transformed to 
a confinement potential of holes.
As in the experiment  \cite{Masaya_double,Masaya_Kondo},
we consider 
the antidot states  formed
by electrons with spin up and down in the lowest Landau level
(i.e.,  the local filling factor is two around the antidot
as in Fig. \ref{fig:experiment}).
We test two kinds of antidot potentials:
bell-shape and inverse parabolic.

\begin{figure}
\includegraphics[width= \columnwidth]{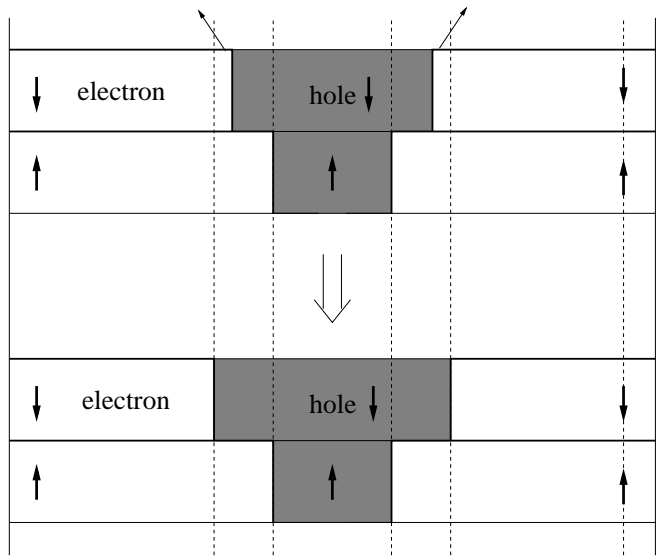}
\caption{\label{fig:normal}
Schematic diagram of particle densities, as a function of distance
from antidot center, in a hole MDD 
where $N_\uparrow$ spin-up holes occupy single-particle states 
with angular momentum $m=0,1, \cdots, N_\uparrow -1$ 
and $N_\downarrow$ spin-down holes 
with $m=0,1, \cdots, N_\downarrow -1$.
Single-particle hole states with spin $\sigma$ are empty 
for $m > N_\sigma - 1$.  
Note that a single-particle state with smaller $m$ is located at 
smaller distance from the center.
The changes in electron and hole densities around the antidot 
are also shown when a spin-down MDD transition 
$| N_\uparrow,N_\downarrow \rangle \to 
| N_\uparrow,N_\downarrow + 1 \rangle$ occurs.
In this process, a spin-down electron tunnels out of the antidot
(indicated as thin arrows) and thus the total hole spin decreases.
}
\end{figure}

For both potentials, the antidot ground states are found to 
be MDDs of holes in certain parameter ranges (see Fig.\ref{fig:normal}).
For a given magnetic field,
the spin configuration of MDD ground states
(the size and the spatial splitting between spin-up and -down 
edges of the droplet)
is determined by the competition \cite{Yang1}
between electron-electron interactions
and the confinement potential of holes:
droplets with larger size are favored by weaker confinement and 
stronger electron-electron interactions.
As magnetic field varies,
the relative magnitude of these two competing factors changes
so that the transitions of MDD ground states can take place.
In general,
there can be three types of transition:
spin-up, spin-down, and spin-flip transitions, which manifest themselves 
via spin-up electron normal resonance, spin-down electron normal resonance, 
and Kondo resonance, respectively.

\begin{figure}
\includegraphics[width= \columnwidth]{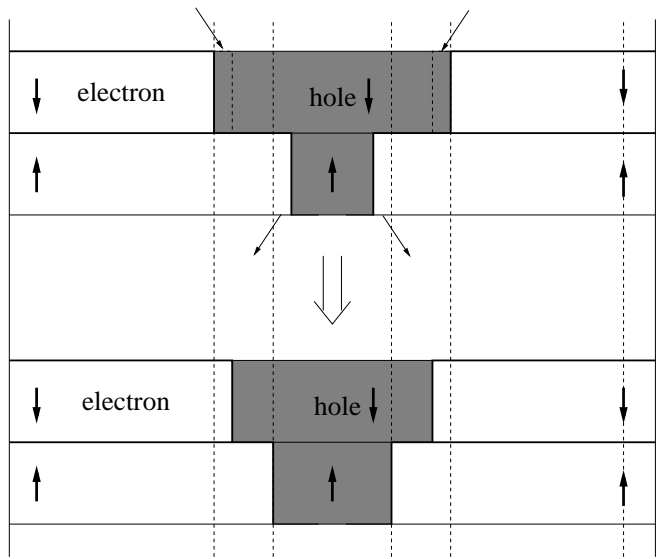}
\caption{\label{fig:kondoflip} 
Same diagram as in Fig. \ref{fig:normal}, but 
for the spin-flip transition 
$| N_\uparrow, N_\downarrow \rangle \to
| N_\uparrow + 1, N_\downarrow - 1 \rangle$.
In this process, a cotunneling event (see thin arrows) takes place,
where a spin-up electron moves out of the antidot while
a spin-down electron moves in,
and thus the total electron (hole) spin decreases (increases).
}
\end{figure}

\begin{figure}
\includegraphics[width= 0.95 \columnwidth]{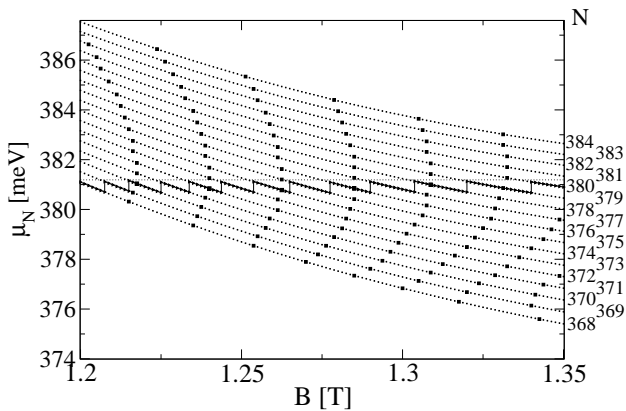}
\caption{\label{fig:chemical2}
Chemical potential, $\mu_N \equiv E_{N+1} - E_N$, 
vs magnetic field $B$ for a bell-shape antidot potential.
Different values of $N = N_\uparrow + N_\downarrow \in [368,384]$ are used.
The horizontal dotted line represents {\it hole Fermi energy},
which can be rather different from electron Fermi energy.
Diamonds represent the spin-flip transition
 $| N_\uparrow, N_\downarrow \rangle \to
| N_\uparrow + 1, N_\downarrow - 1 \rangle$
as $B$ increases,
while vertical jumps show the spin-down transition
 $| N_\uparrow, N_\downarrow \rangle \to
| N_\uparrow, N_\downarrow + 1 \rangle$.
Following the zigzag solid line, the MDD ground state
evolves as
$|N_{\uparrow}, N_{\downarrow} \rangle = |175,193 \rangle 
\rightarrow |175,194 \rangle 
\rightarrow |175,195 \rangle
\Rightarrow |176,194 \rangle 
\rightarrow |176,195 \rangle
\rightarrow |176,196 \rangle
\Rightarrow |177,195 \rangle
\rightarrow |177,196 \rangle
\rightarrow |177,197 \rangle
\Rightarrow |178,196 \rangle
\rightarrow |178,197 \rangle
\rightarrow |178,198 \rangle$
with increasing $B$.
Here $\rightarrow$ and $\Rightarrow $ indicate, respectively, 
the spin-down and spin-flip transitions.
The parameters of this antidot can be found in Sec. VB.
}
\end{figure}

For a bell-shape antidot potential,
we find spin-down (Fig. \ref{fig:normal}) 
and spin-flip transitions (Fig. \ref{fig:kondoflip})
in some ranges of magnetic field, while we do not find spin-up transitions.
In the spin-flip transitions, the number of spin-down (spin-up) holes
decreases (increases) by one as magnetic field becomes stronger. 
A series of these transitions is obtained as a function of magnetic field 
in Fig. \ref{fig:chemical2}.
The features of the transitions are {\it in agreement} with the predictions
of the capacitive interaction model \cite{Sim},
and thus one can explain the experimental observation \cite{Masaya_Kondo}
from them.
From the Hartree-Fock result of the transitions,
we obtain the {\it parameters} of the capacitive interaction model
and study the variation of excess charges as a function of magnetic field.

For an inverse parabolic antidot potential,
we find all the three types of transitions
in some parameter range.
The properties of the spin-flip transition are different 
from those of the bell-shape potential;
in this case, the number of spin-down (spin-up) holes 
increases (decreases) by one as magnetic field increases.
Moreover, the spin-flip transitions are found to appear more frequently 
than the spin-down and spin-up transitions.
As a result, the spatial splitting between spin-up and -down edges of
MDDs becomes larger for stronger magnetic fields.
This behavior
{\it differs} from the experimental situation of Kataoka {\it et al.}, 
where the splitting is expected to be a periodic function of magnetic field.
It would be interesting to investigate inverse parabolic antidots 
experimentally 
and compare the obtained results with our theoretical predictions.

This paper is organized as follows.  
In Sec. II our model Hamiltonian for the antidot is given.   
This Hamiltonian is changed into a hole Hamiltonian
via an electron-hole transformation in Sec. III.
Within a Hartree-Fock approach, we study
the stability of MDDs in Sec. IV and
the MDDs of a bell-shape antidot in Sec. V.
In Sec. VI the properties of the MDDs of the bell-shape antidot 
are shown to be in good agreement with the prediction of 
the capacitive interaction model.
In Sec. VII we investigate the MDDs of an inverse parabolic antidot.
Concluding remarks are given in Sec. VIII.

\section{Model Hamiltonian of antidot}

We consider a 2DEG around an antidot in the presence of 
a strong perpendicular magnetic field $B$ along the z-axis.
Following experiments \cite{Masaya_double,Masaya_Kondo},
the local filling factor around the antidot is chosen to be two.
Thus, the antidot states can be assumed to be formed by 
spin-up and -down electrons in the lowest Landau level.
In the symmetric gauge,
the single electron wave function $\phi_m(r)$
is labeled by the  quantum number $m$, the z-component 
of the angular momentum.  More explicitly, 
$\phi_m(r)=(z/\ell)^m exp(-|z|^2/4\ell^2) / (\sqrt{2^{m+1}\pi m!}\ell)$,
where $z=x+iy$ is the complex coordinate of the 2D plane and
$\ell(B)=\sqrt{\hbar c/e B}= 2.56\times 10^{-6}/\sqrt{B[T]}[cm]$ is the magnetic length. 
We can write the antidot model Hamiltonian:
\begin{eqnarray}
H &=& \sum_\sigma \sum_m^{m_c}W(m) c_{m \sigma}^\dagger c_{m \sigma} 
- \sum_\sigma \sum_m^{m_c} V_m^{ion} c_{m \sigma}^\dagger c_{m \sigma}
\nonumber \\
&-& \frac{1}{2}g \mu B \sum_m^{m_c} 
\left( c_{m \uparrow}^\dagger c_{m \uparrow}
- c_{m \downarrow}^\dagger c_{m \downarrow} \right)
\nonumber \\
&+& \frac{1}{2} \sum_{\sigma_1 \sigma_2} \sum_{m_1' m_2' m_1 m_2}^{m_c}
\left< m_1' m_2' \left| V  \right| m_1  m_2 \right> 
\nonumber \\
&\times& c_{m_1' \sigma_1}^\dagger c_{m_2' \sigma_2}^\dagger c_{m_2 \sigma_2} c_{m_1 \sigma_1},
\label{totalH}
\end{eqnarray}
where $W(m)$ is the antidot potential energy [see Eqs. (\ref{parabolic1},
\ref{bellshape1}) below],
$V$ is the Coulomb interaction, and
$c_{m \sigma}^\dagger$ creates an electron in
the state $\phi_m(r)$ with spin $\sigma$.
The term of $V_m^{ion}$ comes from the {\it neutralizing positive background 
charge} around the antidot. 
Since there are as many ions as the total number of electrons, both spin-up and -down,
we have
$V_m^{ion} = 2 \sum_{m'}^{m_c} \left< m m' |V| m m' \right>$.
By definition $V_m^{ion} > 0 $. 
The angular momentum conservation yields $ {m_1}' + {m_2}' = m_1 + m_2$. 
In our numerical work we use electron states up to a cutoff 
value $m_c$, which is chosen sufficiently large.

The following Coulomb matrix elements \cite{tsi} are used:
\begin{eqnarray}
\langle m+p , n |V|m,n+ p\rangle = 
C^p_{mn}
\left[ A_{m n}^{p} B_{n m}^{p} + B_{m n}^{p} A_{n m}^{p} \right],
\nonumber
\end{eqnarray}
where 
\begin{eqnarray}
A_{m n}^{p} &=& \sum_{i=0}^{m} \binom{m}{i} 
\frac{\Gamma (\frac{1}{2} + i) \Gamma(\frac{1}{2} + p +i)}
{\Gamma (\frac{3}{2} + p + n + i)(p + i)!},
\nonumber \\
B_{m n}^{p} &=& \sum_{i=0}^{m} \binom{m}{i}
\frac{\Gamma (\frac{1}{2} + i) \Gamma(\frac{1}{2} + p +i)}
{\Gamma (\frac{3}{2} + p + n + i)(p + i)!}
(\frac{1}{2} + p + 2 i),
\nonumber \\
C_{m n}^{p} &=& \sqrt{\frac{(m+p)! (n+p)!}{m! n!}} 
\frac{\Gamma (p +m +n + \frac{3}{2})}{\pi 2^{p + m + n +2}}.
\nonumber
\end{eqnarray}
This expression allows us to perform Hartree-Fock calculations 
even when the total number of holes inside antidot is quite large 
(more than 300).

We consider two types of antidot potential in this paper.
The first type is an inverse parabolic potential
\begin{eqnarray}
W(r) = \left\{ \begin{array}{ll}  
\frac{1}{2}\hbar\omega_c-\frac{1}{2}m^{*}\Omega^2r^2, &  r<r_s \\
\textrm{constant}, &    r>r_s. 
\end{array} \right.
\label{parabolic1}
\end{eqnarray}
Beyond $r>r_s$ the potential is flat as  a function of $r$.
The second type is a bell-shape potential
\begin{eqnarray}
W(r) = \left\{ \begin{array}{ll}  \frac{1}{2}\hbar\omega_c-\frac{1}{2}m^{*}\Omega^2r^2, &  r<r_t \\
B+\frac{C}{r^2}, & r_t<r<r_s \\
\textrm{constant}, &    r>r_s.
\end{array} \right.
\label{bellshape1}
\end{eqnarray}
Here $\omega_c = |e| B / (m^{\ast} c)$,
electron charge $e < 0$, and $m^{\ast} = 0.067 m_e$ for GaAs.
In the interval $ r<r_t$ the potential is inverse parabolic, while in the next 
interval $ r_t<r<r_s$ the curvature changes sign.
The matrix elements $W(m)=\langle m|W(r)|m \rangle$ can be 
approximately written as follows when $r_t \gg \ell$ and $r_s-r_t \gg \ell$:
\begin{eqnarray}
W(m) = \left\{ \begin{array}{ll}  \frac{1}{2}\hbar\omega_c-m^*\Omega^2\ell^2(m+1), &  m<m_s \\
\textrm{constant}, &    m>m_s 
\end{array} \right.
\label{parabolic2}
\end{eqnarray}
for the inverse parabolic potential, while
\begin{eqnarray}
W(m) = \left\{ \begin{array}{ll}  \frac{1}{2}\hbar\omega_c-m^*\Omega^2\ell^2(m+1), &  m<m_t \\
W_0 + \frac{C}{2\ell^2m}, & m_t<m<m_s \\
\textrm{constant}, &   m >m_s 
\end{array} \right.
\label{bellshape2}
\end{eqnarray}
for the bell-shape potential. 
Here $m_i=r_i^2/(2\ell^2)-1$,
where $i=s,t$.
In order to make sure that the bell-shape potential is 
continuous at $m=m_t$, we set 
$W_0 = \hbar\omega_c/2 - m^*\Omega^2\ell^2(m_t +1) - C/(2\ell^2m_t)$.
Note that $W(m)$ is monotonously decreasing with $m$.

We remark that the values of  $r_t$ and $r_s$ are fixed by 
the shape of $W(r)$, while $m_t$ and $m_s$ are
magnetic-field dependent. 
But we can use constant $m_t$ and $m_s$ 
since, for the narrow range of magnetic field ($\delta B \approx 0.02$ T) 
of interest in this work,
the change $\delta m_i \approx m_i \delta B/(2B)$
is of the order of one, much smaller than $m_t$ and $m_s$.
In our bell-shape antidot,  
both positions of spin-up and -down edges of MDDs
are in the interval of $r_t<r<r_s$
so that they feel $1/r^2$ potential.
Note that  $m_s$ is different from $m_c$, 
which is the cutoff for the single particle states.
In our numerical work we choose $m_c=400$ and $m_s=300$. 
The typical value of $m_t$ is between 100 and 200.

\section{Electron-hole transformation}

An electron antidot system is an open geometry problem and often requires
heavy calculations to compute its physical properties.
Such a difficulty can be avoided by transforming
an electron antidot system to a system which confines holes 
since the transformed system contains only a finite number of holes.
Such transformation is described in this section.

We consider a particle-hole transformation 
\cite{antidot_PhysicaE,Johnson}
of the type 
$c_{m,\sigma} \rightarrow h_{m,\sigma}^\dagger$ 
and $c_{m,\sigma}^\dagger \rightarrow h_{m,\sigma}$. 
The term representing the interaction with the positive background 
in Eq. (\ref{totalH}) transforms into
\begin{eqnarray}
-\sum_\sigma \sum_m^{m_c} V_m^{ion} c_{m \sigma} ^\dagger c_{m \sigma} 
= 2 \sum_\sigma \sum_m^{m_c} V_m^H 
h_{m \sigma}^\dagger h_{m \sigma} -4 \sum_m^{m_c} V_m^H,
\nonumber
\end{eqnarray}
where
$V_m^H=\sum_{m'}^{m_c} \langle m m' |V| m m' \rangle$.
The total Hamiltonian in Eq. (\ref{totalH}) can be rewritten as
\begin{eqnarray}
H &=& \sum_m^{m_c} \varepsilon_m h_{m \uparrow}^\dagger h_{m \uparrow} 
+ \sum_m^{m_c} \varepsilon_m h_{m \downarrow}^\dagger h_{m \downarrow}
\nonumber \\
&+& \frac{1}{2} g \mu B \sum_m^{m_c} 
\left( h_{m \uparrow}^\dagger h_{m \uparrow}
- h_{m \downarrow}^\dagger h_{m \downarrow} \right) 
\nonumber \\
&+& \frac{1}{2} \sum_{\sigma_1 \sigma_2} \sum_{{m_1}' {m_2}' m_1 m_2}^{m_c} 
\langle m_1' m_2' |V|m_1 m_2 \rangle 
\nonumber \\
&\times& h_{m_2 \sigma_2}^\dagger h_{m_1 \sigma_1}^\dagger 
h_{{m_1}' \sigma_1} h_{{m_2}' \sigma_2 } 
\nonumber \\
&+& 2 \sum_m^{m_c} W(m) -2 \sum_m^{m_c} V_m^H - \sum_m^{m_c} V_m^X,
\label{holeHam}
\end{eqnarray}
where 
$V_m^X = \sum_{m'}^{m_c} \langle m m' |V| m' m \rangle$.
The effective single hole energy is 
\begin{eqnarray}
\varepsilon_m = - W(m) + V_m^X.
\label{single_hole_energy}
\end{eqnarray}
The first term of Eq. (\ref{single_hole_energy})
is the confinement energy coming from $W(m)$ after the transformation, 
while the second term represents the
change in exchange energy when an electron with angular momentum $m$ disappears.
A Hartree term is absent in $\varepsilon_m$, since it is canceled by 
the interaction between a hole with the quantum number $m$ and 
the positive background charge. 
So in this transformed Hamiltonian of hole systems
the background charge term is {\it absent}.
Note that the Zeeman term in Eq. (\ref{holeHam}) has the opposite sign 
to the corresponding term in Eq. (\ref{totalH}). 

There is a simple check of this result.
If there are zero holes, according to Eq. (\ref{holeHam}),
the total energy is
$E_T = 2 \sum_m^{m_c} W(m) -2 \sum_m^{m_c} V_m^H - \sum_m^{m_c} V_m^X$.
This energy should be equal to the total energy of an electron antidot 
system in which electrons are occupied 
from 0 to $m_c$;
in this case the Hartree-Fock theory exactly gives the total energy, 
which has four contributions of the confinement energy
($2 \sum_m^{m_c} W(m)$),
interaction energy with the positive background
($-4 \sum_m^{m_c} V_m^H$),
Hartree ($2 \sum_m^{m_c} V_m^H$),
and exchange energies ($-\sum_m^{m_c} V_m^X$).
The constant terms ($E_T$) in the hamiltonian of Eq. (\ref{holeHam})
will be ignored hereafter.

When the cutoff value $m_c$ is larger than the total number of holes,
it is a good approximation to treat $V_m^X$ 
in Eq. (\ref{single_hole_energy})
as a constant \cite{Yang2}.
Then, $\varepsilon_m$ is a monotonously increasing function of $m$.
For the bell-shape potential, for example, 
$d\varepsilon_m/dm \simeq -dW(m)/dm = C/(2m^2\ell^2) > 0$ in the interval $m_t<m<m_s$
[see Eq. (\ref{bellshape2})].

Note that the magnetic field dependence of the hole confinement 
potential is
\begin{eqnarray}
-W(m) \sim 
\left\{ \begin{array}{ll} 1/B, &  \textrm{for parabolic} \\
 B, & \textrm{for bell-shape ($\sim 1/r^2$ region)}.
\end{array} \right.
\label{Bdependence}
\end{eqnarray}
On the other hand, the interaction energy scale $e^2/(\epsilon \ell)$ is 
proportional to $\sqrt{B}$.  Therefore, the hole confinement 
potential relative to the interaction energy, $-W(m)/[e^2/(\epsilon \ell)]$, 
decreases with increasing $B$ for an inverse parabolic antidot, 
while it increases in the $1/r^2$ potential region of a bell-shape antidot.

\begin{figure}
\includegraphics[width=  \columnwidth]{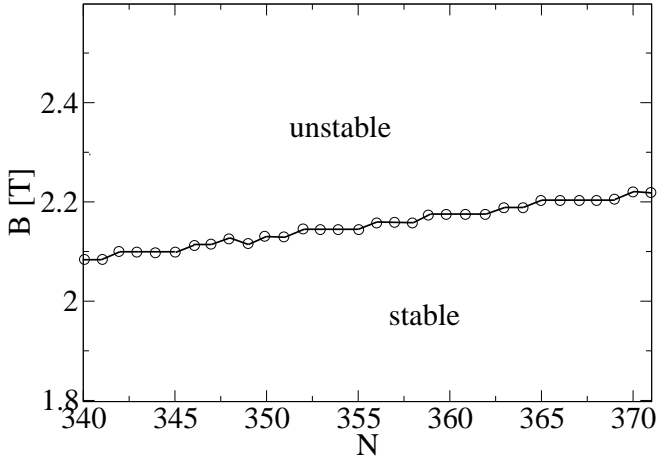}
\caption{\label{fig:phase}
The boundary (critical magnetic fields) of stable and unstable MDDs of a bell-shape
antidot. The following parameters are used: $\hbar\Omega=1.5$ meV, $C/(2\ell^2)=2100\sqrt{B\ (\mathrm{Tesla})}\times e^2/(\epsilon \ell)$, and $m_t=118$.  
}
\end{figure}

\section{Hole maximum density droplets}
In electron quantum dots, MDDs are exact ground states
when a strong magnetic field is applied and the confinement potential
is strong enough. 
The reason is as follows: 
A MDD is an eigenstate of $L_z$, the $z$-component of 
the total angular momentum.  In fact, it has the smallest 
possible eigenvalue of $L_z$ for a given number of electrons 
and there are no other states in the Hilbert space with 
the same eigenvalue.  
If one chooses a rotationally symmetric potential, 
a MDD is also an eigenstate of the Hamiltonian.  
Since the mean radius of the single electron wave packet 
increases with the $z$-component of the angular momentum, 
the confinement potential favors small values of $L_z$.  
Therefore, a MDD is certainly {\em the} exact ground state if the 
confinement potential is infinitely strong.  
It must also remain so in a certain parameter range as 
long as the potential is strong enough.
The properties of a MDD were investigated by exact 
diagonalization for a small dot \cite{Yang1}.
For a dot with about 50 electrons,
the properties of spin-polarized MDDs
and their instability were also studied \cite{Yang2,Yang3}
using Hartree-Fock approach and exact diagonalization.
Experimental investigation of electron MDDs were reported
by several groups \cite{Oosterkamp}.

Our antidot problem becomes similar to the quantum dot case 
after the particle-hole transformation described in the last section.
Like electron MDDs,  hole MDDs have a single-Slater-determinant form,
\begin{eqnarray}
| N_\uparrow, N_\downarrow \rangle = 
h_{{N_\downarrow -1},\downarrow}^\dagger \cdots h_{0 \downarrow}^\dagger
h_{{N_\uparrow -1},\uparrow}^\dagger \cdots h_{0 \uparrow}^\dagger 
| 0 \rangle.
\end{eqnarray}
The total number of holes is $N = N_\uparrow + N_\downarrow$, 
and $N_\downarrow$ is equal to or larger than $N_\uparrow$ 
due to the Zeeman energy.

The hole MDD is excellently described by the Hartree-Fock approach.
Its total Hartree-Fock energy is
\begin{eqnarray}
E_{HF}( N_\uparrow, N_\downarrow ) = E_H + E_X + E_Z + E_C,
\label{HFenergy}
\end{eqnarray}
where $E_H$, $E_X$, and $E_Z$ are the Hartree, exchange,
and Zeeman energies, respectively.
Here,
\begin{eqnarray}
E_H &=& 
\frac{1}{2} \sum_m^{N_\uparrow -1} \sum_{m'}^{N_\uparrow -1} V_{m m'}^H 
+ \frac{1}{2} \sum_m^{N_\downarrow -1} \sum_{m'}^{N_\downarrow -1} V_{m m'}^H
\nonumber \\
&+& \sum_m^{N_\downarrow -1} \sum_{m'}^{N_\uparrow -1} V_{m m'}^H,
\nonumber \\
E_X &=& - \frac{1}{2} 
\sum_m^{N_\uparrow -1} \sum_{m'}^{N_\uparrow -1} V_{m m'}^X
- \frac{1}{2} \sum_m^{N_\downarrow -1} \sum_{m'}^{N_\downarrow -1} V_{m m'}^X,
\nonumber \\
E_Z &=& \frac{1}{2} g \mu B \left( N_\uparrow - N_\downarrow \right),
\nonumber \\
E_C &=& \sum_m^{N_\uparrow -1} \left( -W(m) + V_m^X \right)
+ \sum_m^{N_\downarrow -1} \left( -W(m) + V_m^X \right).
\nonumber
\end{eqnarray}
In the above expressions, we have used the definitions 
$ V_{m m'}^H = \langle m m' |V| m m'\rangle$ and 
$ V_{m m'}^X = \langle m m' |V| m' m\rangle$.
        
\begin{figure}
\includegraphics[width= \columnwidth]{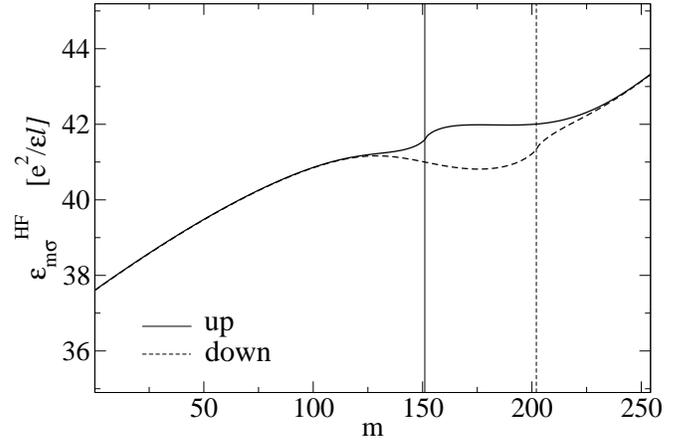}
\caption{\label{fig:stable}
The renormalized Hartree-Fock single hole energies 
$\varepsilon_{m \sigma}^{HF}$ of a MDD are plotted 
for both spin up (solid) and down (dashed line).
This MDD is stable as
$\varepsilon_{m \sigma}^{HF}\leq \varepsilon_{N_{\sigma}-1,\sigma}^{HF}$
for all $m\leq N_{\sigma}-1$ and for both spins.
We choose parameters  $\hbar\Omega=1.5$ meV, $C/(2\ell^2)=2100\sqrt{B\ (\mathrm{Tesla})}\times e^2/(\epsilon \ell)$, $m_t=118$, $N=355$, 
and $B=2.13$T.
The last occupied states ($m=N_\sigma - 1$) are marked with vertical lines. 
Note that $N_\uparrow=152$ and $N_\downarrow=203$.
}
\end{figure}

\begin{figure}
\includegraphics[width=  \columnwidth]{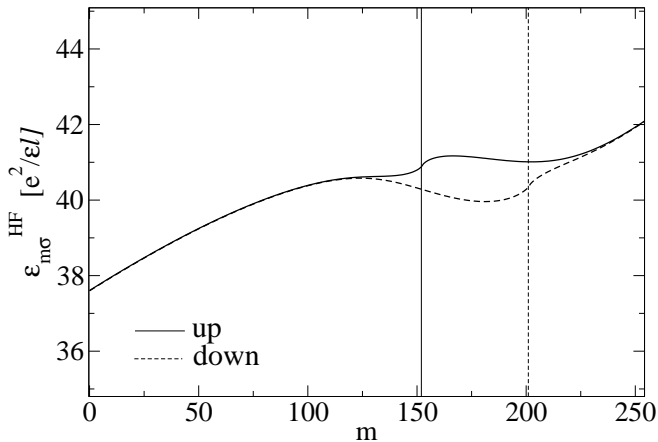}
\caption{\label{fig:unstable} 
Same curves as in Fig. \ref{fig:stable} but at $B=2.205$T.
This MDD is unstable since  
$\varepsilon_{m \sigma}^{HF} > \varepsilon_{N_{\sigma}-1,\sigma}^{HF}$ 
for  $m\approx 125$.
Note that $N_\uparrow=153$ and $N_\downarrow=202$.
}
\end{figure}

From Eq. (\ref{HFenergy}),
one can define the renormalized single hole energy,
which includes the Hartree and exchange self-energy corrections 
\begin{eqnarray}
\varepsilon_{m \sigma}^{HF} &=& -W(m) + V_m^X 
+ \sum_{m'}^{N_\sigma -1} V_{m m'}^H
\nonumber \\
&+& \sum_{m'}^{N_{\overline{\sigma}} -1} V_{m m'}^H  
- \sum_{m'}^{N_\sigma -1} V_{m m'}^X 
+ g \mu B s_\sigma,
\end{eqnarray}
where $s_\uparrow = 1/2$ and $s_\downarrow = -1/2$.  
This renormalized single hole energy is useful for studying
the stability of MDD states.
A MDD state will be a stable ground state if the
occupied single hole states satisfy 
$\varepsilon_{m \sigma}^{HF} \leq 
\varepsilon_{(N_{\sigma}-1)\sigma}^{HF}$ 
for all $m\leq N_{\sigma}-1$ and for both spins.
For a given $N$, this condition is satisfied, i.e., MDD states are stable,
below some critical magnetic fields.
A phase boundary between stable and unstable states is displayed in 
Fig. \ref{fig:phase}.
In calculating the phase boundary for a given value of $N$,
we calculate the ground configuration $| N_\uparrow, N_\downarrow \rangle$
and $\varepsilon_{m \sigma}^{HF}$ with varying $B$, 
and determine the critical value of $B$, where the above 
mentioned stability condition is violated.
The examples of stable and unstable MDDs are plotted in
Figs. \ref{fig:stable} and \ref{fig:unstable}, respectively.

The physics determining the spin configuration 
$(N_\uparrow,N_\downarrow)$ of a MDD ground state is as follows:
There is a competition \cite{Yang1} between the Coulomb energy ($E_H + E_X$) 
and the confinement energy ($E_C$).
If the confinement energy is strong, the total energy is minimized 
by making the confinement energy small, i.e., by making the droplet 
size small. 
As a result, for a given $N$, 
the configuration with smaller $N_{\downarrow}-N_{\uparrow}$ 
is favored by stronger confinement, since the droplet size is
determined by $N_\downarrow$.
On the other hand,
if the confinement energy is weak,
the total energy can be minimized by making the Coulomb energy smaller,
i.e., by making the droplet size larger.
Thus the degree of spatial splitting
between spin-up and -down edges
(or $N_{\downarrow}-N_{\uparrow}$) of MDDs
depends on the relative strength of the confinement energy and 
electron-electron interaction.

\section{Hartree-Fock results of bell-shape antidots}

In this section, we study the bell-shape antidot with the potential
of Eq. (\ref{bellshape1}).
Before discussing the Hartree-Fock result of its hole ground states,
it is instructive to consider possible transitions of the hole
ground states.  These transitions  can occur since the competition between
the Coulomb energy and the confinement energy of MDDs varies with
magnetic field.
They correspond to resonant tunneling processes 
\cite{Sim} of the antidot
when the antidot states are weakly coupled to
extended edge channels (see Fig. \ref{fig:experiment}).

\subsection{possible ground state transitions}

In general, there exist three kinds of transitions
between MDD ground states, 
which are spin-up, spin-down, and spin-flip transitions.
Within the Hartree-Fock approximation, the spin-up transition of
$|N_\uparrow, N_\downarrow \rangle \to
|N_\uparrow \pm 1, N_\downarrow \rangle$
occurs at the magnetic fields where the degeneracy
\begin{eqnarray}
E_{HF}(N_\uparrow, N_\downarrow) = 
E_{HF}(N_\uparrow \pm 1, N_\downarrow) 
\label{hole_resonance_condition_up} 
\end{eqnarray}
is satisfied.
The plus and minus signs refer to tunneling in and out of a hole,
respectively.
Similarly,
the spin-down transition occurs when
\begin{eqnarray}
E_{HF}(N_\uparrow, N_\downarrow) = 
E_{HF}(N_\uparrow, N_\downarrow \pm 1).
\label{hole_resonance_condition_down}
\end{eqnarray}
In electron language, the spin-up (-down) transitions correspond to
normal resonant tunnelings of spin-up (-down) electrons \cite{Sim}.
The normal spin-down resonant tunneling, for example,
is illustrated in Fig. \ref{fig:normal}.
With increasing magnetic field, 
the direction of these transitions is
either
$|N_\uparrow, N_\downarrow \rangle \to
|N_\uparrow + 1, N_\downarrow \rangle$
or
$|N_\uparrow, N_\downarrow \rangle \to
|N_\uparrow, N_\downarrow + 1 \rangle$
so that the total number ($N = N_\uparrow + N_\downarrow$) of holes
becomes larger.
This increase of the total hole number
can be understood from the fact that
the effective hole potential $-W(m)$ decreases with 
magnetic field while the Fermi energy does not change.
On the other hand, the spin-flip transitions occur at the magnetic
field where the degeneracy
\begin{equation}
E_{HF}(N_\uparrow, N_\downarrow) =
E_{HF}(N_\uparrow \pm 1, N_\downarrow \mp 1)
\label{hole_resonance_condition_Kondo} 
\end{equation}
is satisfied.
These processes can cause \cite{Sim} the Kondo resonance
\cite{Kondo,Hewson}
in the antidot system when both the spin-up and -down
parts of MDDs are weakly coupled to the corresponding extended
edge channels;
this resonance is accompanied by the cotunneling processes where
an electron (hole) tunnels into antidot
and another electron (hole) with the opposite spin tunnels out 
via a virtual state
$|N_\uparrow \pm 1, N_\downarrow \rangle$ or
$|N_\uparrow, N_\downarrow \mp 1 \rangle$.

We note that as magnetic field increases, the spin-flip transitions
occur as either
$| N_\uparrow, N_\downarrow \rangle \to
| N_\uparrow + 1, N_\downarrow - 1 \rangle$
(see Fig. \ref{fig:kondoflip})
or
$| N_\uparrow, N_\downarrow \rangle \to
| N_\uparrow - 1, N_\downarrow + 1 \rangle$,
depending on the detailed shape of antidot potential such as 
its curvature near MDD edges.
For the bell-shape antidot (see Sec. VB), we find the former direction of
transition, while the latter for the inverse parabolic antidot (Sec. VII).

\subsection{Hartree-Fock results}
In this subsection, we perform Hartree-Fock calculations for
the bell-shape antidot with more than 300 holes and study
the transition of MDD ground states.
We find the spin-flip and spin-down transitions 
in certain parameter ranges.

\begin{figure}
\includegraphics[width=0.95 \columnwidth]{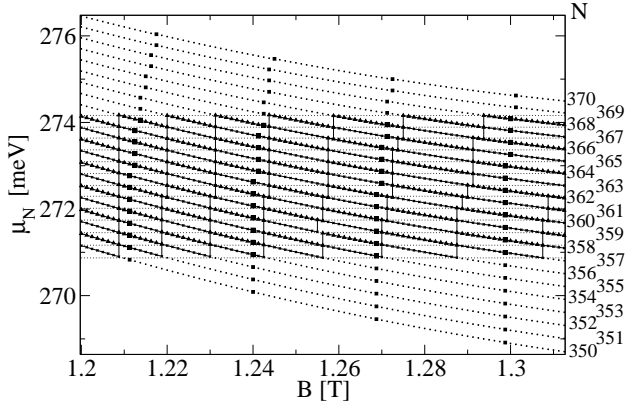}
\caption{\label{fig:chemical} 
Same curves as in Fig. \ref{fig:chemical2} but for
a weaker bell-shape potential.
Twenty one different values of $N \in [350,370]$ are used.  
Following the topmost zigzag solid line, for example, the MDD ground state 
evolves as
$|N_{\uparrow}, N_{\downarrow} \rangle = |128,233 \rangle 
\rightarrow |128,234 \rangle 
\Rightarrow |129,233 \rangle
\rightarrow |129,234 \rangle 
\rightarrow |129,235 \rangle
\Rightarrow |130,234 \rangle
\rightarrow |130,235 \rangle
\rightarrow |130,236 \rangle
\Rightarrow |131,235 \rangle
\rightarrow |131,236 \rangle
\rightarrow |131,237 \rangle
\Rightarrow |132,236 \rangle$ 
with increasing $B$.
Similarly, in the fourth zigzag solid line, 
$|N_{\uparrow}, N_{\downarrow} \rangle$ 
evolves as
$|127,231 \rangle 
\rightarrow |127,232 \rangle 
\Rightarrow |128,231 \rangle 
\rightarrow |128,232 \rangle 
\rightarrow |128,233 \rangle 
\Rightarrow |129,232 \rangle 
\rightarrow |129,233 \rangle 
\rightarrow |129,234 \rangle 
\Rightarrow |130,233 \rangle$. 
Note that both of $N_\uparrow$ and $N_\downarrow$ 
increase with $B$.
}
\end{figure}

From the Hamiltonian of Eq. (\ref{HFenergy}),
we find the energy of MDD ground state $| N_\uparrow, N_\downarrow \rangle$
with varying magnetic field $B$
for a given total number of holes $N=N_{\uparrow}+N_{\downarrow}$.
In Fig. \ref{fig:chemical}, the chemical potential
$\mu_N \equiv E_{N+1} - E_N$,
which is the energy difference of ground states,
is plotted as a function of $B$.
In this calculation, we use 
$\hbar \Omega=1.5$ meV, $C/(2\ell^2)=2100\sqrt{B\ (\mathrm{Tesla})}\times e^2/(\epsilon \ell)$, 
$N \in [350,370]$, and $m_t = 118$.
The Fermi energies of holes are shown as 
the horizontal dotted lines in Fig. \ref{fig:chemical}.
For the selected parameters and magnetic field ranges,
the bell-shape antidot is found to have the properties 
that 
(i) MDD ground states satisfy $N_\downarrow > N_\uparrow > m_t$,
(ii) only the spin-down and spin-flip transitions appear,
and (iii) both the transitions are periodic with $B$
(the periods of spin-down and spin-flip transitions are
$\Delta B_\downarrow = 0.0113$ T and 
$\Delta B_K = 0.0288$ T, respectively);
there are no spin-up transition.  
Figure \ref{fig:chemical2} displays the chemical potential 
for a stronger confinement potential, whose parameters 
are $\hbar \Omega=1.9$ meV, $C/(2\ell^2)=3600\sqrt{B\ (\mathrm{Tesla})}\times e^2/(\epsilon \ell)$,
$N \in [368,384]$, and $m_t = 111$. 
The differences between $N_\uparrow$ and $N_\downarrow$ are smaller 
than those in Fig. \ref{fig:chemical}.

\begin{figure}
\includegraphics[width= \columnwidth]{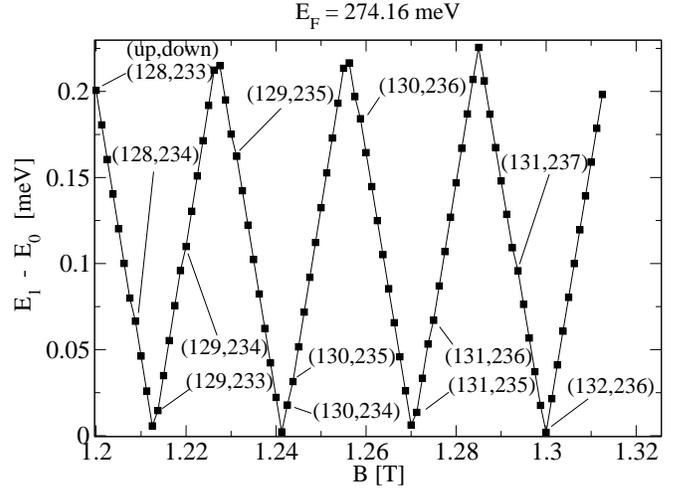}
\caption{\label{fig:epsart} 
Energy differences of the ground and first excited states of the antidot
studied in Fig. \ref{fig:chemical} as a function of $B$.
We choose the topmost Fermi level shown in Fig. \ref{fig:chemical} so that 
the transition of MDD ground states (their spin configurations are marked 
in this figure) follows the top zigzag line of Fig. \ref{fig:chemical}.
At a cusp in $E_1(B) - E_0(B)$, the spin-down transition 
$|N_\uparrow, N_\downarrow \rangle \to
|N_\uparrow, N_\downarrow + 1 \rangle$ occurs.
The spin-flip transitions 
$| N_\uparrow, N_\downarrow \rangle \to
| N_\uparrow + 1 ,N_\downarrow - 1 \rangle$ are also found around
$B=1.21375, 1.2425, 1.27125 $, and $1.3 \mathrm{T}$. 
A domain boundary between two MDD ground states is indicated 
by the numbers $(N_\uparrow,N_\downarrow)$. 
The state right to the domain boundary has $N_\uparrow$ and $N_\downarrow$ 
numbers of spin-up and -down holes, respectively.
}
 \end{figure}

In Fig. \ref{fig:epsart}, we calculate the spin configuration 
$(N_\uparrow, N_\downarrow)$ of MDD ground states and 
the energy difference between the ground and first excited states 
as a function of $B$. The topmost Fermi level of holes  
in Fig. \ref{fig:chemical} is chosen for these calculations.
In this figure,
the system starts in 
$|N_\uparrow, N_\downarrow \rangle = |128, 233 \rangle$ 
and changes into $|128, 234 \rangle$ 
at about $0.01$ T larger magnetic field.
This spin-down transition 
$| N_{\uparrow},N_{\downarrow} \rangle \to
| N_{\uparrow},N_{\downarrow}+1 \rangle$
corresponds to 
the chemical potential jumps in Fig. \ref{fig:chemical}
occurring whenever each chemical potential line intersects the Fermi level
(see the changes in the occupation numbers of spin-up and -down
electrons in Fig. \ref{fig:normal}).
As $B$ increases further, 
there appears a spin-flip transition of
$|128, 234 \rangle \to |129, 233 \rangle$
around $B=1.215$ T
(see the corresponding degeneracy point marked by a square dot
in Fig. \ref{fig:chemical} and the changes in the occupation numbers
of spin-up and -down electrons in Fig. \ref{fig:kondoflip}).
For the studied bell-shape antidots, we find 
only the spin-flip transition of the type
$|N_\uparrow, N_\downarrow \rangle \to
|N_\uparrow + 1, N_\downarrow - 1 \rangle$
with increasing $B$,
where the total hole (electron) spin increases (decreases).
This can be understood from the fact that
for the bell-shape antidot with  a negative 
potential curvature  near MDD edges ( $N_\uparrow, N_\downarrow > m_t$),
the confinement potential behaves like 
$\sim B$
for the states with $m > m_t$
[see Eq. (\ref{Bdependence})],
while the Coulomb matrix elements goes like $\sim \sqrt{B}$.
Consequently, the total confinement energy of holes
increases faster than the total Coulomb energy as $B$ increases.
This causes the hole droplet to minimize its size, i.e.,
$N_{\downarrow}$ becomes smaller for stronger magnetic field.
This feature is absent in the inverse parabolic antidot,
where the spin-flip transition is found to occur as
$| N_\uparrow, N_\downarrow \rangle \to
| N_\uparrow - 1, N_\downarrow + 1 \rangle$
with increasing $B$
(see Sec. VII).

However, the spatial splitting 
($\sim \sqrt{N_\downarrow} \ell - \sqrt{N_\uparrow} \ell$)
of spin-up and -down edges of MDDs is much larger than $\ell$
in the bell-shape antidot studied in Fig. \ref{fig:chemical}.
In this case the coupling of the spin-up part of the MDDs to the extended
edge channels will be negligibly small, compared to that of the spin-down
part, and thus, the Kondo resonance can not occur at the spin-flip
transition points.
Instead, these parameters may represent  a good model of
$(h/2e)$ Aharonov-Bohm oscillations without Kondo resonances
measured in Ref. \cite{Masaya_double}.
To see the Kondo resonance, one should have the conditions \cite{Sim}
that (i) there exists the spin-flip transitions
and (ii) both the spin-up and spin-down electrons are 
weakly coupled to extended edge channels;
the condition (ii) is equivalent to 
$R_\downarrow - R_\uparrow \lesssim \ell$ in MDD states,
where
$R_\uparrow \sim \sqrt{2N_\uparrow} \ell$ and 
$R_\downarrow \sim \sqrt{2N_\downarrow} \ell$
are 
the radii of spin-up and -down droplets, respectively.
It is possible to make 
$R_\downarrow - R_\uparrow$ smaller 
by making the confinement potential stronger.
Then, the corresponding MDD 
would be closer to the actual electronic state 
measured in Ref. \cite{Masaya_Kondo}.
Figure \ref{fig:chemical2} displays  
such a case with smaller $R_\downarrow - R_\uparrow$.

As will be shown in the next section VI,
the spin-down and spin-flip transitions found in the bell-shape
antidots are in good agreement with the predictions of the capacitive 
interaction model \cite{Sim},
and thus they can explain qualitatively the experimental conductance 
data \cite{Masaya_Kondo}
of Aharonov-Bohm oscillations with Kondo-like signatures.
However, for realistic quantitative comparison,
the coupling to higher Landau levels by electron-electron
interactions may have to be included.

\section{Capacitive interaction model of bell-shape antidots}

Both the spin-down and spin-flip transitions 
shown in Figs. \ref{fig:chemical2} and \ref{fig:chemical} 
are almost periodic with $B$,
indicating that the bell-shape antidot with more than 300 holes 
may be large enough to be described by the phenomenological capacitive 
interaction model for antidot \cite{Sim}.
Our Hartree-Fock results are indeed consistent with the model.
Below this is demonstrated by determining the parameters of the model
from the Hartree-Fock results of the antidot studied 
in Fig. \ref{fig:chemical}.
For the antidot in Fig. \ref{fig:chemical2} one can get the parameters
in the same way.

\subsection{Excess charges}

The capacitive interaction model was developed for the case that
the local filling factor is two around antidot.
In this model, the excess charges \cite{Sim}
can be defined as 
\begin{equation}
\delta q_\sigma (N_\sigma, B) = e (N_\sigma - \tilde{N}_\sigma(B))
\label{excess_charge} 
\end{equation}
in terms of the number of holes $N_\sigma$ and $\sigma = \uparrow, 
\downarrow$.
The function $\tilde{N}_\sigma(B)$ 
has the meaning of ``the optimal number'' of spin-$\sigma$ holes 
that minimizes the total energy in the absence of holes 
with the opposite spin.  Note that $\tilde{N}_\sigma(B)$ is real-valued 
while the actual number of holes is an integer, which prohibits 
continuous change of $N_\sigma$ and leads to Coulomb blockade.  
Of course, the detailed form of $\tilde{N}_\sigma(B)$ 
depends on the shape of the antidot potential near its edge.  
For a sufficiently small range of $B$, 
one can use an approximately  linear form 
\begin{equation}
e \tilde{N}_\sigma(B) = e (a_\sigma B + b_\sigma).
\label{excess_charge2} 
\end{equation}
Here, $a_\sigma e$ is the rate of excess charge accumulation with 
increasing $B$ and $b_\sigma e$ originates  from the 
positive background charge. 
Both $a_\sigma$ and $b_\sigma$ are taken as constants.
This is a good approximation for large-size antidots in strong
magnetic fields.
Then, one has
\begin{eqnarray}
\frac{\delta q_\sigma (N_\sigma, B)}{e}=N_{\sigma}-a_\sigma B-b_\sigma. 
\label{excess_charge3}
\end{eqnarray}

By assuming that the excess charges interact capacitively,
one can write the total energy of an isolate antidot as \cite{Sim}
\begin{eqnarray}
E_\textrm{CI}(\delta q_{\uparrow},\delta q_{\downarrow})
&=& \frac{1}{2}\sum_{\sigma,\sigma'}
\delta q_{\sigma}(C^{-1})_{\sigma,\sigma'}\delta q_{\sigma'} \nonumber\\
&=&\frac{(\delta q_{\downarrow}+\alpha\delta q_{\uparrow})^2}{2C_{out}}
+ \frac{\delta q_{\uparrow}^2}{2C_{in}},
\label{CIenergy}
\end{eqnarray}
where 
$C$ is a capacitive matrix,
$\alpha=|C_{\uparrow\downarrow}|/C_{\uparrow\uparrow}$,
$C_{out}=C_{\downarrow\downarrow}-\alpha|C_{\uparrow\downarrow}|$,
and $C_{in}=C_{\uparrow\uparrow}$.
These elements of capacitive matrix can be taken as constants
for large-size antidot with $\Delta B_{AB} \ll B$, where
$C_{\sigma,\sigma'}$ can vary very slowly over several 
Aharonov-Bohm periods $\Delta B_{AB}$.
Thus,
for large-size antidots in strong magnetic fields,
the capacitive interaction model defined by 
Eqs. (\ref{excess_charge3},\ref{CIenergy})
is a good approximation.
The model is 
determined by the constant
parameters 
$\alpha$, $C_{in}$, $C_{out}$, $a_\sigma$, and $b_\sigma$,
and is analogous to the constant interaction model
of quantum dots \cite{Glazman_CI}.

\subsection{Transition conditions}

The conditions of transitions between MDD ground states can be
rewritten by using $E_\textrm{CI}$.
For example, the condition (\ref{hole_resonance_condition_down})
for the spin-down transition of
$|N_\uparrow,N_\downarrow \rangle \to
|N_\uparrow,N_\downarrow \pm 1 \rangle$
becomes
$E_\mathrm{CI}(\delta q_\uparrow, \delta q_\downarrow \pm e) = 
E_\mathrm{CI}(\delta q_\uparrow, \delta q_\downarrow)$,
which is equivalent to
\begin{eqnarray}
\frac{\delta q_\downarrow(N_\downarrow,B)}{e}\pm\frac{1}{2}+
\alpha\frac{\delta q_\uparrow(N_\uparrow,B)}{e}=0.
\label{CI_resonance_condition_down}
\end{eqnarray}
It is useful to note that 
$N_\downarrow$ increases (decreases) with increasing (decreasing) $B$
in the spin-down transitions, as discussed in Sec. VA.
This observation makes it easy to choose signs in 
Eq. (\ref{CI_resonance_condition_down}).

The spin-flip transition in 
Eq. (\ref{hole_resonance_condition_Kondo}) 
can be written as
$E_\mathrm{CI}(\delta q_\uparrow \pm e, \delta q_\downarrow \mp e) = 
E_\mathrm{CI}(\delta q_\uparrow, \delta q_\downarrow)$.
When the transition 
$|N_\uparrow,N_\downarrow \rangle \to
|N_\uparrow \pm 1, N_\downarrow \mp 1 \rangle$
appears at $B$,
one can rewrites this condition as
\begin{eqnarray}
\frac{\alpha - 1}{C_{out}} 
\left(\delta q_\downarrow(N_\downarrow \mp \frac{1}{2},B)
+ \alpha \delta q_\uparrow(N_\uparrow \pm \frac{1}{2},B)\right)
\nonumber \\
+ \frac{1}{C_{in}} \delta q_\uparrow(N_\uparrow \pm \frac{1}{2},B) = 0.
\label{CI_resonance_condition_Kondo} 
\end{eqnarray}
In the spin flip transition,
$N_\downarrow$ decreases (increases) with increasing (decreasing) $B$.
Again this observation makes it easy to choose signs 
in Eq. (\ref{CI_resonance_condition_Kondo}).
When  $\alpha = 1$,
the spin-flip condition of Eq. (\ref{CI_resonance_condition_Kondo})
is reduced to a simple form 
\begin{eqnarray}
\frac{\delta q_\uparrow}{e}=
\pm\frac{1}{2}.
\label{CI_resonance_condition_Kondo2}
\end{eqnarray}

\subsection{Evolution of excess charges}

As $B$ varies, excess charges are accumulated following 
Eq. (\ref{excess_charge3}).  The  
pair of excess charges
$(\delta q_\uparrow/e,\delta q_\downarrow/e)$ thus evolves
with satisfying  
\begin{eqnarray}
\frac{\delta q_\downarrow}{e}=
\frac{\delta q_\uparrow}{e}
\frac{a_\downarrow}{a_\uparrow}
+\beta,
\label{excess_charge6}
\end{eqnarray}
where 
$\beta = 
N_\uparrow+(b_\uparrow-N_\downarrow)(a_\downarrow/a_\uparrow)
-b_\downarrow$. This condition describes a line in the space of $(\delta q_\uparrow/e,\delta q_\downarrow/e)$.
If the magnetic field $B$ is tuned such that one of 
the spin-up, spin-down, and spin-flip transition conditions
in Eqs. (\ref{hole_resonance_condition_up},
\ref{hole_resonance_condition_down}, 
\ref{hole_resonance_condition_Kondo}) is satisfied,
$(N_\uparrow,N_\downarrow)$ and $(N'_\uparrow,N'_\downarrow)$ 
become degenerate.
At this degenerate point the excess charges $\delta q_\sigma(N_\sigma,B)$ jump to
their transition values $\delta q_\sigma(N'_\sigma,B)$.
As a result, the evolution of the excess-charge pair
$(\delta q_\uparrow/e,\delta q_\downarrow/e)$
is restricted
within a hexagonal cell
whose boundaries are determined by
the six transition conditions 
in Eqs. (\ref{hole_resonance_condition_up},
\ref{hole_resonance_condition_down}, 
\ref{hole_resonance_condition_Kondo}) \cite{Sim}.
Note that the transition conditions can be easily rewritten
in terms of $\delta q_\sigma$ as in Eqs. 
(\ref{CI_resonance_condition_down},
\ref{CI_resonance_condition_Kondo}).
The evolution trajectory of the excess-charge pair in the cell
can be simply drawn when the value of the pair at starting $B$
is known:
The trajectory evolves along a line parallel to
$\delta q_\downarrow = (a_\downarrow/a_\uparrow) \delta q_\uparrow$
and jumps to the corresponding opposite boundary when
it collides with a cell boundary.

\subsection{Parameters of capacitive interaction model}

One can obtain the parameters such as
$\alpha$, $a_\sigma$, and $b_\sigma$ of the capacitive interaction 
model from the Hartree-Fock results of the $B$-dependence of
MDD ground states $| N_\uparrow, N_\downarrow \rangle$.
Below, we show such a procedure for a ground-state
transition on the chemical potential line which
has the value $\mu = 273.33$ meV 
in Fig. \ref{fig:chemical}.

For this case, one can find the evolution sequence of 
$|N_\uparrow,N_\downarrow \rangle$ as a function of $B$ 
in Fig. \ref{fig:chemical}
(see the fourth zigzag line from the top one as well as the caption).
From the periodic spin-down and spin-flip transitions one can first
determine $\alpha$ and $a_\sigma$.
Applying  Eq. (\ref{CI_resonance_condition_down}) to 
three consecutive spin-down transitions in $B \in (1.205 ,1.245)$ T,
we get $\alpha = 1$ and $a_\uparrow + a_\downarrow = 88.89$.
Similarly, applying Eq. (\ref{CI_resonance_condition_Kondo2}) to 
two consecutive spin-flip transitions at $B = 1.2125$ T and 1.24125 T,
we find $a_\uparrow = 34.78$ and $a_\downarrow=54.11$.
For $\alpha=1$, it is noteworthy that
the spin-flip condition in 
Eq. (\ref{CI_resonance_condition_Kondo}) can be simply reduced to
Eq. (\ref{CI_resonance_condition_Kondo2}),
and that the evolution trajectory of
$(\delta q_\uparrow,\delta q_\downarrow)$ is restricted
in a parallelogram, instead of a hexagonal cell,
since no spin-up transition appears. 

The other parameters $b_\sigma$'s can be uniquely determined from
the assumption that the evolution speed (the accumulation speed of
excess charges) is constant with varying $B$
and from the facts that (i) the trajectory moves within the cell
and (ii)
at the spin-flip transitions (e.g., at $B=1.2125 $T in this case)
the value of $\delta q_\uparrow$ is fixed by Eq. 
(\ref{CI_resonance_condition_Kondo2}) when $\alpha = 1$.
Combining these with  Eq. (\ref{excess_charge3}) 
we obtain $b_\uparrow = 85.33$ and $b_\downarrow = 165.79$.
Note that 
one can not determine $C_{out}$ and $C_{in}$ 
from the data in Fig. \ref{fig:chemical} 
when $\alpha = 1$.

\begin{figure}
\includegraphics[width= \columnwidth]{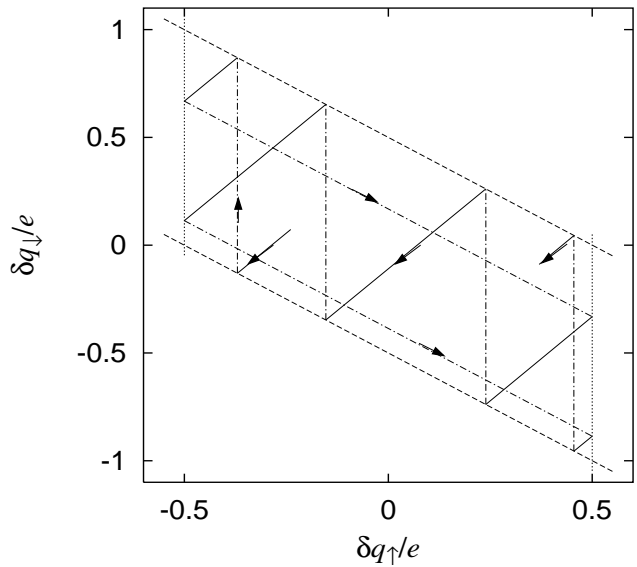}
\caption{\label{fig:HFtrajectory} 
The evolution trajectory (solid line) of 
$(\delta q_\uparrow(B), \delta q_\downarrow(B))$ 
for the bell-shape antidot potential, while the MDD ground state
evolves following the fourth zigzag solid line in
Fig. \ref{fig:chemical}.
The conditions of
the spin-down [Eq. (\ref{hole_resonance_condition_down})]
and spin-flip [Eq. (\ref{hole_resonance_condition_Kondo})]
transitions define the cell boundaries
(dashed and dotted lines, respectively).
The trajectory is determined by
Eq. (\ref{excess_charge2}), thus it
evolves parallel to the line of
$\delta q_\downarrow = (a_\downarrow / a_\uparrow) \delta q_\uparrow$.
When the trajectory collides with a cell boundary,
it jumps following vertical (spin-down) or diagonal dashed-dot
lines (spin-flip transition).
The trajectory starts at $B = 1.205$ T and ends at 1.245 T, and
the arrows represent the direction of evolution with
increasing $B$.
The parameters ($\alpha = 1$, $a_\downarrow / a_\uparrow = 1.55$,
$b_\uparrow = 85.33$, $b_\downarrow = 165.79$) are determined from
the result in Fig. \ref{fig:chemical} (see text).
}
\end{figure}

\subsection{Evolution trajectory of excess charge}

We draw the evolution trajectory 
$(\delta q_\uparrow,\delta q_\downarrow)$
by using Eq. (\ref{excess_charge6}) and
the parameters of $\alpha$, $a_\sigma$, and $b_\sigma$
obtained in the last subsection VID.
The evolution trajectory  describes well the features of 
the ground state transitions in the range $B \in (1.205 ,1.245)$ T,
such as the sequence of transitions
and the differences of magnetic fields between adjacent transitions.
In the range of $B \gtrsim 1.25$ T, the trajectory becomes to deviate
from the Hartree-Fock results in Fig. \ref{fig:chemical}, indicating that
the parameters are not constant but slowly vary with $B$.
This finding demonstrates that the capacitive interaction model
can describe the behavior of the bell-shape antidot very well.

Some predictions of the  capacitive interaction model are  useful 
in  understanding of microscopic 
numerical results.
It predicts that $\alpha = 1$ when
the interaction between
$\delta q_\uparrow$ and $\delta q_\downarrow$ is maximal,
indicating that
the spin-up and -down excess charges of MDD states formed in
the bell-shape antidot are strongly coupled.
It also predicts that
when $\alpha = 1$,
no spin-up transition appears due to Coulomb blockade
while
periodic spin-down and spin-flip transitions with $B$ can occur
(the periods are $\Delta B_\downarrow$ and $\Delta B_K$, 
respectively).
When $\alpha = 1$, the ratio of the transition periods
relates to the ratio of accumulation speeds of excess charges
as 
$\Delta B_K / \Delta B_\downarrow = 1 + a_\downarrow / a_\uparrow$.

The ratio of accumulation speed $a_\downarrow/a_\uparrow$ of 
excess charges
generally relates to the ratio of hole occupation area 
$R_\downarrow^2 / R_\uparrow^2$, since larger area gives
smaller Aharonov-Bohm period and thus faster accumulation of excess charge.
Our system has the area ratio
$R_\downarrow^2 / R_\uparrow^2 = N_\downarrow / N_\uparrow \simeq 1.8$
when $(N_\uparrow, N_\downarrow) = (128, 233)$, while
the speed ratio is $a_\downarrow / a_\uparrow = 1.56$.
The discrepancy between
$a_\downarrow / a_\uparrow$ and
$R_\downarrow^2 / R_\uparrow^2$
originates from the fact that
the accumulation speed depends on antidot potential shape. The
ground states of the tested antidot
have large spatial splitting between the edges of spin-up and spin-down 
states so that the outmost spin-up orbital at $m=N_\uparrow$ sees 
a different potential slope from the spin-down orbital at $m=N_\downarrow$.
For the case of $N_\uparrow \simeq N_\downarrow$
we expect
$a_\downarrow / a_\uparrow \simeq R_\downarrow^2 / R_\uparrow^2$.

When $a_\downarrow / a_\uparrow \simeq 1$ 
(i.e., $N_\uparrow \simeq N_\downarrow$),
one gets $\Delta B_{\rm K} \simeq 2 \Delta B_\downarrow$
from the relation of
$\Delta B_K / \Delta B_\downarrow = 1 + a_\downarrow / a_\uparrow$.
Then, one has two spin-down and one spin-flip transitions in one
Aharonov-Bohm
period, which is the type (i) evolution trajectory of excess charges
predicted in Ref. \cite{Sim}.
The type (i) trajectory matches very well 
with the experimental findings of
$(h/2e)$ AB oscillation and Kondo-like signatures
in Ref. \cite{Masaya_Kondo}.

\begin{figure}
\includegraphics[width=  \columnwidth]{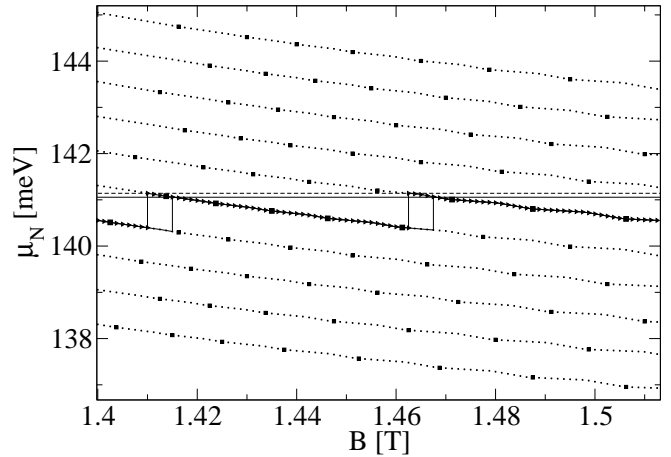}
\caption{\label{fig:parabolic}
Magnetic field dependence of the chemical potential for an 
inverse parabolic antidot with $\hbar\Omega=1.5$ meV.
Chemical potentials are shown for two different values of hole Fermi energies.
Squares indicate spin-flip transitions, while vertical jumps
show spin-up or spin-down transitions.
In the spin-flip transitions,
the occupation number configuration changes from the lower to
the upper panels of Fig. \ref{fig:kondoflip} as $B$ increases.
For the Fermi level drawn by the dashed horizontal line,
$| N_\uparrow, N_\downarrow \rangle$ evolves as
$| 66, 67 \rangle \Rightarrow
|65,68 \rangle \rightarrow 
|66,68 \rangle \Rightarrow 
|65,69 \rangle \Rightarrow 
|64,70 \rangle \Rightarrow 
|63,71 \rangle \Rightarrow 
|62,72 \rangle$ 
as $B$ increases from 1.4 T.
Here $\Rightarrow$ indicates the spin-flip transition, while
$\rightarrow$ shows the spin-up or -down transition.
For the Fermi level drawn by the solid  horizontal line,
$| N_\uparrow, N_\downarrow \rangle$ evolves as
$| 66, 67 \rangle \Rightarrow
|65,68 \rangle \rightarrow 
|65,69 \rangle \Rightarrow 
|64,70 \rangle \Rightarrow 
|63,71 \rangle \Rightarrow 
|62,72 \rangle$. 
The main different  result between the upper and lower Fermi energies is that the vertical jumps
represent the spin-up and -down transitions, respectively.}
\end{figure}

\section{Hartree-Fock results of inverse parabolic antidots}

In this section, we consider the inverse parabolic antidot
\cite{antidot_PhysicaE}
shown in Eqs. (\ref{parabolic1},\ref{parabolic2}).
From the Hartree-Fock calculation, 
the ground state is found to be a MDD
when we choose the parameters 
$\hbar\Omega=1.5$ meV and $B \sim 1.45$ T.
The transition of MDD ground states in this antidot
is plotted in Fig. \ref{fig:parabolic}.
Three types of transitions can appear.
As $B$ increases,
while the direction of the spin-up (-down) transition is
$| N_\uparrow ,N_\downarrow \rangle  \to 
| N_\uparrow + 1, N_\downarrow \rangle$
($| N_\uparrow ,N_\downarrow \rangle  \to 
| N_\uparrow ,N_\downarrow + 1 \rangle$),
the spin-flip transition occurs as
$| N_\uparrow ,N_\downarrow \rangle \to
| N_\uparrow - 1,N_\downarrow + 1 \rangle$.  Note that
the direction of the spin-flip transition 
is opposite to the bell-shape case.
This can be understood from the fact that
$\sim 1/B$
in the inverse parabolic case
[see Eq. (\ref{Bdependence})],
while Coulomb matrix elements goes like $\sim \sqrt{B}$.
Due to this magnetic field dependence, the arguments based on the total energy minimization 
indicates that the hole droplet can have a larger size
(larger $N_\downarrow$) for a stronger $B$.
The features of spin-flip transition around an antidot thus
depend on its potential shape.
The Kondo effect associated with similar spin-flip transitions in a parabolic 
electron quantum dot is studied theoretically in Ref. \cite{choi}.

In the case of the inverse parabolic antidot,
the spin-flip transitions cause that
$N_\downarrow$ becomes larger with $B$.
And, they appear more frequently with varying $B$ than the
spin-up and spin-down transitions.
As a result, the hole spin polarization
($N_\uparrow - N_\downarrow$) decreases as $B$ increases.
In electron language, this means that the electron spin polarization
around the antidot increases with $B$.
These features cannot be explained by the predictions of the capacitive
interaction model \cite{Sim} and do not match with the experimental data of Kataoka et al
\cite{Masaya_Kondo},
which indicate that the spin polarization around the antidot is 
periodic with $B$.
It would be interesting to investigate the properties 
of an inverse parabolic antidot experimentally.

\section{Concluding remarks}

We have investigated electronic properties of antidots 
in the integer quantum Hall regime
(the local filling factor around the antidot is two),
by using an electron-hole transformation and a Hartree-Fock approach.
Our numerical work shows that
when the antidot potential is strong enough,
hole MDDs of a single-Slater-determinant form are stable in a  certain
parameter range and represent exactly the ground states.

For a bell-shape antidot with more than 300 holes,
we find that
there exist the spin-down and spin-flip transitions between MDD ground states
as magnetic field varies,
and that their properties, such as sequence of transitions and the
spin polarization of MDDs, are in good agreement with the
phenomenological capacitive interaction model.
Thus, hole MDDs and the bell-shape antidot could be a good model for
the antidot studied experimentally by Kataoka {\it et al.}
The Hartree-Fock results of the ground-state transitions allow us
to obtain
the parameters of the capacitive interaction model
and the accumulation of excess charges around the antidot
as a function of magnetic field.

The properties of hole MDDs depend on the competition between
electron-electron interactions and hole confinement potential. 
The competition determines
the degree of spin splitting between spin-up and -down edges of MDD:
stronger hole confinement (weaker Coulomb energy) favors
the droplets with smaller size.
As a consequence of this competition 
the direction of the spin-flip transitions 
depends on the detailed potential shape 
(e.g., the potential curvature near MDD edges).
For example,
in the bell-shape antidot
the spin-flip transition occurs as
$| N_\uparrow ,N_\downarrow \rangle \to
| N_\uparrow + 1,N_\downarrow - 1 \rangle$
with increasing $B$,
while
$| N_\uparrow ,N_\downarrow \rangle \to
| N_\uparrow - 1,N_\downarrow + 1 \rangle$
in the inverse parabolic antidot.
Thus, the physical properties of antidots are connected to the shape of 
the antidot potential. 
It would be interesting to investigate antidots with different shapes 
and compare their properties with each other.

For a better quantitative comparison with the experimental data  the coupling to higher Landau levels
by the Coulomb scattering 
may have to be included.
However, it is a challenge to calculate the matrix elements of 
the screened Coulomb interaction analytically.
It is also desirable to use a more smoothly varying antidot potential 
than the one we used here 
(the first derivative our potential as a function of $m$ 
changes suddenly at $m_t$).   

Finally, reconstruction of the edge structure
around the antidot may take place, similar to the extended edge channels
\cite{Chklovskii,Chamon,Oaknin}.
The electron-hole transformation makes it possible to
investigate numerically the edge reconstruction near the antidot, which
requires a proper treatment of correlation effects 
beyond Hartree-Fock approximation\cite{Yang3,Yang4}.\\

\begin{acknowledgements}
We thank M. Kataoka for useful discussions.
We were supported by QSRC at Dongguk University (SREY and HSS) and
SKORE-A (HY). 
\end{acknowledgements}


\end{document}